\documentclass{article}

\usepackage{arxiv}

\usepackage[utf8]{inputenc} 
\usepackage[T1]{fontenc}    
\usepackage{hyperref}       
\usepackage{url}            
\usepackage{booktabs}       
\usepackage{amsfonts}       
\usepackage{nicefrac}       
\usepackage{microtype}      
\usepackage{lipsum}

\usepackage{amsmath,amssymb}
\usepackage{hyperref}
\usepackage{graphicx}
\usepackage{stfloats}
\usepackage{bm}
\usepackage{textgreek}
\usepackage{url}
\usepackage{enumitem}
\usepackage{comment}
\usepackage{multirow}
\usepackage{cite}

\newcommand{\beq}{\begin{equation}}
\newcommand{\eeq}{\end{equation}}
\newcommand{\bea}{\begin{eqnarray}}
\newcommand{\eea}{\end{eqnarray}}
\newcommand{\invivo}{\textit{in~vivo}}
\newcommand{\fdg}{${}^{18}\mathrm{F}$-FDG }

\title{Super-resolution PET imaging using convolutional neural networks}

\author{
  Tzu-An~Song\\
  Department of Electrical and Computer Engineering\\
  University of Massachusetts Lowell\\
  Lowell, MA 01854 \\
     \And
 Samadrita R.~Chowdhury \\
  Department of Electrical and Computer Engineering\\
  University of Massachusetts Lowell\\
  Lowell, MA 01854 \\
   \And
 Fan~Yang \\
  Department of Electrical and Computer Engineering\\
  University of Massachusetts Lowell\\
  Lowell, MA 01854 \\
\And
 Joyita~Dutta \\
  Department of Electrical and Computer Engineering\\
  University of Massachusetts Lowell\\
  Lowell, MA 01854 \\
  Gordon Center for Medical Imaging\\
  Massachusetts General Hospital\\
  Boston, MA 01720 \\
}

\begin{document}
\maketitle

\begin{abstract}
Positron emission tomography (PET) suffers from severe resolution limitations which limit its quantitative accuracy. In this paper, we present a super-resolution (SR) imaging technique for PET based on convolutional neural networks (CNNs). To facilitate the resolution recovery process, we incorporate high-resolution (HR) anatomical information based on magnetic resonance (MR) imaging. We introduce the spatial location information of the input image patches as additional CNN inputs to accommodate the spatially-variant nature of the blur kernels in PET. We compared the performance of shallow (3-layer) and very deep (20-layer) CNNs with various combinations of the following inputs: low-resolution (LR) PET, radial locations, axial locations, and HR MR. To validate the CNN architectures, we performed both realistic simulation studies using the BrainWeb digital phantom and clinical neuroimaging data analysis. For both simulation and clinical studies, the LR PET images were based on the Siemens HR+ scanner. Two different scenarios were examined in simulation: one where the target HR image is the ground-truth phantom image and another where the target HR image is based on the Siemens HRRT scanner --- a high-resolution dedicated brain PET scanner. The latter scenario was also examined using clinical neuroimaging datasets. A number of factors affected relative performance of the different CNN designs examined, including network depth, target image quality, and the resemblance between the target and anatomical images. In general, however, all CNNs outperformed classical penalized deconvolution techniques by large margins both qualitatively (e.g., edge and contrast recovery) and quantitatively (as indicated by two metrics: peak signal-to-noise-ratio and structural similarity index).  
\end{abstract}

\keywords{Super-resolution \and CNN \and Deep learning \and PET \and MRI \and Multimodality imaging \and Partial volume correction}

\section{Introduction}
Positron emission tomography (PET) is a 3D medical imaging modality that allows $\invivo$ quantitation of molecular targets. While oncology \cite{Sotoudeh_jmri} and neurology \cite{Catana_jnm} are perhaps the fields where PET is of the greatest relevance, its applications are expanding to many other clinical domains \cite{Salata2017role,Hess2016pet}. The quantitative capabilities of PET are confounded by a number of degrading factors, the most prominent of which are low signal-to-noise ratio (SNR) and intrinsically limited spatial resolution. While the former is largely driven by tracer dose and detector sensitivity, the latter is the driven by a number of factors, including both physical and hardware-limited constraints and software issues. The resolution limitations pose an even greater challenge when the target regions-of-interest (ROIs) are smaller. Physical and hardware-related factors limiting the spatial resolution include the non-collinearity of the emitted photon pairs, intercrystal scatter, crystal penetration, and the non-zero positron range of PET radionuclides \cite{leahy2000stat,dutta2013quant,dutta2013non}. On the software front, resolution reductions are largely a product of smoothing regularizers and filters commonly used within or post-reconstruction for lowering the noise levels in the final images \cite{qi2000res}. Together image blurring and tissue fractioning (due to spatial sampling for image digitization) lead to the so-called \textit{partial volume effect} that is embodied by spillover of estimated activity across different ROIs \cite{rousset1998}. 

Broadly, the efforts to address the resolution challenge encompass both within-reconstruction and post-reconstruction corrections. The former family includes methods that incorporate image-domain or sinogram-domain point spread functions (PSFs) in the PET image reconstruction framework \cite{reader2003algorithm,alessio2006,panin2006} and/or smoothing penalties that preserve edges by incorporating anatomical information \cite{leahy1991,comtat2002,bowsher2004utilizing,baete2004,bataille2007brain,pedemonte20114D,somayajula2011} or other transform-domain information \cite{wang2012pen,kim2015dynamic,wang2015edge}. The latter family of post-reconstruction filtering techniques includes both non-iterative corrections \cite{meltzer1990,mullergartner1992,rousset1998,soret2007,thomas2011,bousse2012} and techniques that rely on an iterative deconvolution backbone \cite{vancittert1931,richardson1972bayesian,lucy1974iterative} which is stabilized by different edge-guided or anatomically-guided penalty or prior functions \cite{yan2015,Song2019pet}.

Unlike partial volume correction, strategies for which are often modality-specific, super-resolution (SR) imaging is a more general problem in image processing and computer vision.  SR imaging refers to the task of converting a low-resolution (LR) image to a high-resolution (HR) one. The problem is inherently ill-posed as there multiple HR images that may correspond to any given LR image. Classical approaches for SR imaging involve collating multiple LR images with subpixel shifts and applying motion-estimation techniques for combining them into an HR image frame \cite{Nasrollahi2012super,Wallach2012super}. Less computationally-intensive modern approaches to SR encompass the family of so-called ``example-based" techniques \cite{Glasner2009,Kim2010SingleImageSR,Yang2010image,Timofte2013,Yang2013fast,Jia2013,Schulter2013}, which exploit self-similarities or recurring redundancies within the same image by searching for similar ``patches" or sub-images within a given image. These methods are particularly effective for super-resolving natural images with fine and repetitive textures and less meaningful for most medical image types. With the proliferation of deep learning techniques, many deep SR models have been proposed and have demonstrated state-of-the-art performance at SR tasks. Deep SR models commenced with a paper that proposed a 3-layer CNN architecture (commonly referred to in literature as the SRCNN) \cite{Dong2016image}. Subsequently, a very deep SR (VDSR) CNN architecture \cite{Kim2016accurate} that had 20 layers and used residual learning \cite{He2009deep} was demonstrated to result in much-improved performance over the shallower SRCNN approach.  More recently SR performance has been further boosted by leveraging generative adversarial networks (GANs) \cite{Ledig2017photo}, although GAN training remains notoriously difficult. Our previous work on SR PET spans a wide gamut, including both (classical) penalized deconvolution based on joint entropy \cite{Dutta2015pet,Song2019pet} and deep learning approach based on the VDSR CNN \cite{Song2018super}. This work is an extension of the latter effort. 

In this paper, we design, implement, and validate several CNN architectures for SR PET imaging, including both shallow and very deep varieties. As a key adaptation of these models to PET imaging, we supplement the LR PET input image with its HR anatomical counterpart, e.g. a T1-weighted MR image. Unlike uniformly-blurred natural images, PET images are blurred in a spatially-variant manner \cite{alessio2006,cloquet2010}. We, therefore, further adjust the network to accommodate spatial location details as inputs to assist the SR process. ``Ground-truth" HR images are required for the training phase of all supervised learning methods. Though simulation studies are not constrained by this demand, it is a challenge for clinical studies where it is usually infeasible to obtain the ``ground-truth" HR counterparts for LR PET scans of human subjects. To ensure clinical utility of this method, we extend it to training based on imperfect target images derived from a higher-resolution clinical scanner, thereby exploiting the SR framework to establish a mapping from an LR scanner's image domain to an HR scanner's image domain. In section \ref{sec:theory} of this paper, we present the underlying network architecture. In section \ref{sec:methods}, we describe the simulation data generation steps and the network training and validation procedures. In section \ref{sec:results}, we present simulation results comparing the performance of CNN-based SR with two well-studied reference approaches. A summary of our work and a discussion on the application of CNNs for SR PET imaging are presented in section \ref{sec:conclusion}.

\section{Theory}
\label{sec:theory}
\subsection{Network Design}
\subsubsection{CNN Basics}
CNNs typically contain three types of layers: convolutional layers, nonlinear activation layers, and pooling layers. Pooling layers, which ensure shift invariance, are useful for classification and object recognition tasks. For SR imaging, which is an estimation/regression task, it suffices for the CNN to contain only convolutional and activation layers. A convolutional layer contains an array of convolutional kernels which extracts linear features from an input based on local connectivity. The pixel intensity at location $(i,j)$ in the $k$th feature map for the $l$th layer, $z^{lk}_{ij}$, can be mathematically represented as \cite{Gu2018RecentAI}:
\beq
z^{lk}_{ij}={\bm{w}^{lk}}^T\bm{x}^{l}_{ij}+b^{lk},     
\eeq
where $\bm{x}^{l}_{ij}$ is the input patch at the $(i,j)$th pixel location and $\bm{w}^{lk}$ and $b^{lk}$ are the kernel weights and bias respectively. An activation layer introduces nonlinearities that enable the extraction of nonlinear features by the CNN. In deep neural networks, the rectified linear unit (ReLU) is a popular choice for the activation function as, unlike alternatives such as the sigmoid function, it does not exhibit the vanishing gradient problem \cite{LeCun1998EfficientB}. The ReLU activation function is defined as:
\beq
\bm{z}=\max(\bm{x},\bm{0}),
\eeq
where $\bm{x}$ is the input vector and $\bm{z}$ is the output vector. In addition, a ReLU activation function can accelerate computation of a network model because ReLUs set input values that are negative to be zero. 

\subsubsection{Residual learning}
The first CNN applied to SR imaging (SRCNN) \cite{Dong2016image} sought to directly estimate the SR image as the network output. In contrast to such methods which compute the latent clean image, we adopt the residual learning strategy to remove the latent clean image from the blurry observations. This strategy was originally described in the \textit{ResNet} \cite{He2009deep} architecture for image recognition and has proven particularly beneficial for very deep networks, for which training accuracy degrades with increasing network depth. A generalization of this idea based on \textit{residual blocks} demonstrated even higher efficacy at solving the SR problem \cite{Lim2017EnhancedDR}.

\begin{figure}[!t]
\centering
\includegraphics[width=0.4\linewidth]{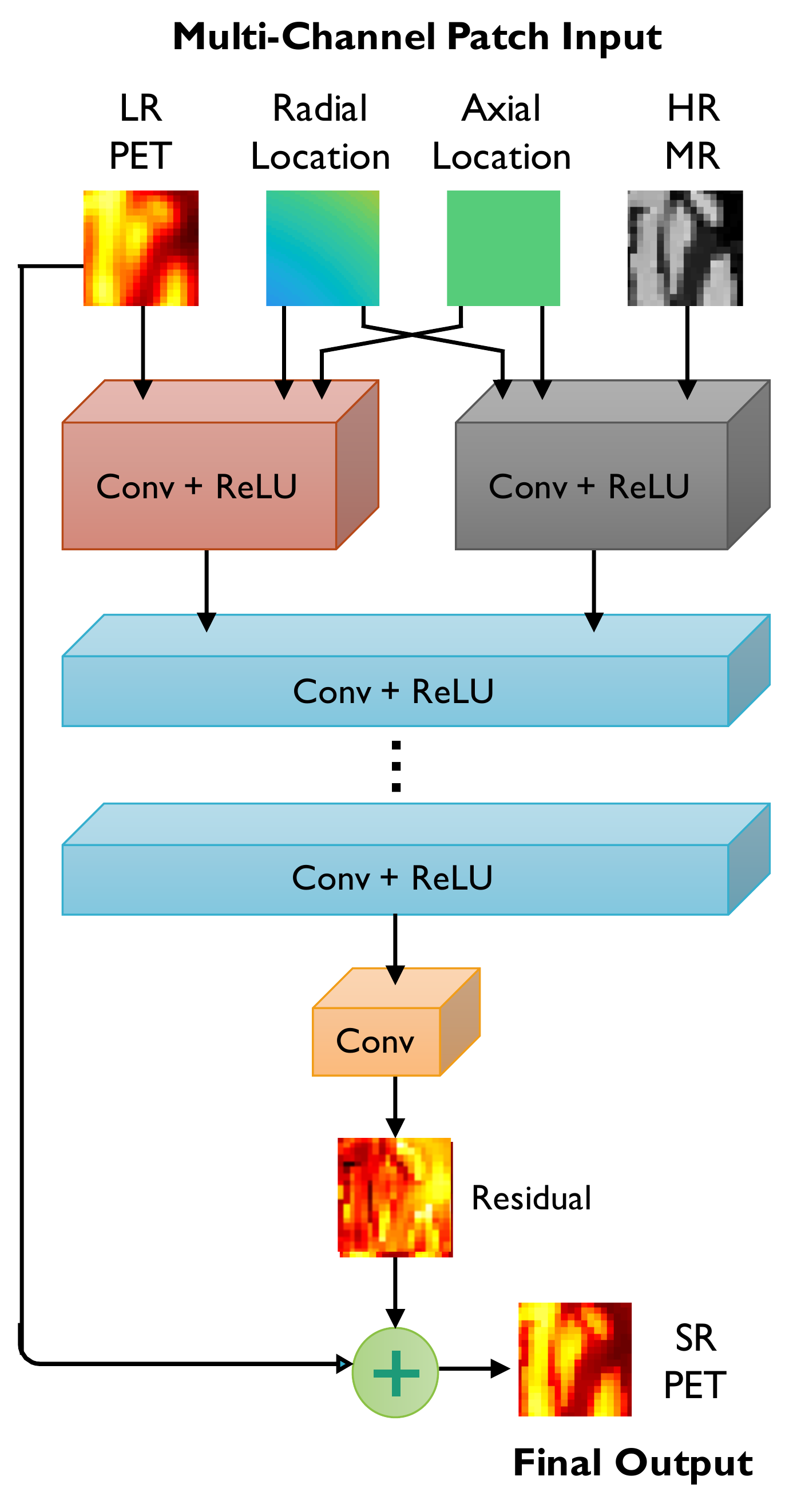}
\caption{CNN architecture for SR PET. The network uses up to 4 inputs: (i) LR PET (the main input), (ii) HR MR, (iii) radial locations, and (iv) axial locations. The networks include alternating convolutional (Conv) and nonlinear activation (ReLU) layers. It predicts the residual PET image which is later added to the input LR PET to generate the SR PET.}
\label{fig:network}
\end{figure}

\subsection{Network Inputs}
\label{ssec:inputs}
One key contribution of this paper is the tailoring of the CNN inputs to address the needs of SR PET imaging. All CNNs implemented here have the LR PET as the main input. This is similar to the two papers that inspired this work \cite{Dong2016image,Kim2016accurate}, both of which had the LR single- or multi-channel images as their inputs. To further assist the resolution recovery process, additional inputs are incorporated as described below:
\subsubsection{Anatomical Inputs}
In generating the SR PET images, we seek to exploit the similarities between the PET and its high-resolution anatomical counterpart. Most clinical and preclinical PET scanners come equipped with anatomical imaging capabilities, in the form of computed tomography (CT) or MR imaging, to complement the functional information in PET with structural information. As illustrated in the schematic in Fig. \ref{fig:network}, we employ CNNs with multi-channel inputs, that include LR PET and HR MR input channels. 
\subsubsection{Spatial Inputs}
We provide location details to the network via additional input channels in the form of patches representing radial and axial coordinate locations. In light of the cylindrical symmetry of PET scanners, we deem these sufficient for learning the spatially-variant structure of the blurring operator directly from the training data. 
\subsubsection{Fusion}
For natural RGB images, where the three input channels typically have a high degree of structural similarity, usually the same set of kernels are effective for feature extraction. In contrast, our CNN architecture exhibits a higher degree of input heterogeneity. Our initial experiments indicated the need for greater network width to accommodate a diverse set of features based on the very different input channels. We found an efficient solution to this by using separate kernels at the lower levels and fusing this information at higher levels as demonstrated in Fig. \ref{fig:network}.
\subsection{Network Depth}
A key finding of the VDSR paper was that the deeper the network, the better its performance. The paper showed that VDSR outperforms SRCNN by a great margin in terms of PSNR \cite{Kim2016accurate}. Here, we implement networks with different depths:
\subsubsection{Shallow CNN}
We designed and implemented a set of ``shallow" networks loosely inspired by the 3-layer SRCNN \cite{Dong2016image}.  The modifications used here are 1) residual learning and 2) modified inputs as described in section \ref{ssec:inputs}. It only has three convolutional layers, each followed by a ReLU, except for the last output layer. While, as a rule of thumb, the qualifier ``deep" is applied to networks with three or more layers, we use the word ``shallow" in this paper in a relative way. 
\subsubsection{Very Deep CNN}
We designed and implemented a series of ``very deep" CNNs based on the VDSR architecture in \cite{Kim2016accurate}, but with a different input design described in section \ref{ssec:inputs}. The very deep SR networks all have 20 convolutional layers, each followed by a ReLU, except for the last output layer.

\begin{table}[!ht]
\caption{CNN architectures}
\label{tab:cnn}
\centering
\footnotesize
\begin{tabular}{c|c|c|c|c|c|c|c|c}
\hline
\hline
Network & S1 & V1 & S2 & V2 & S3 & V3 & S4 & V4 \\
properties &  &  &  &  &  &  &  &  \\
\hline
\hline
Number & 3 & 20 & 3 & 20 & 3 & 20 & 3 & 20\\
of layers &  &  &  &  &  &  &  & \\
\hline
Input types & LR PET & LR PET & LR PET & LR PET & LR PET & LR PET & LR PET & LR PET\\
 & & & HR MR & HR MR & Radial & Radial & HR MR & HR MR\\
 	& & & & & locations & locations & & \\
 & & & & & Axial & Axial & Radial & Radial \\
 	& & & & &  locations &  locations &  locations &  locations\\
 & & & & & & & Axial  & Axial \\
  	& & & & & & &  locations &  locations\\
\hline
\end{tabular}
\end{table}

\subsection{Network Types}
We implemented, validated, and compared both shallow (3-layer) and  very deep (20-layer) CNN architectures with varying numbers of inputs. For the rest of the paper, we refer to these configurations as S1, S2, S3, S4, V1, V2, V3, and V4 as summarized in Table \ref{tab:cnn}. 

\section{Methods}
\label{sec:methods}
\subsection{Overview} 
\label{sssec:over}
In the following, we describe two simulation studies using the BrainWeb digital phantom and the \fdg radiotracer and a clinical patient study also based on ${}^{18}\mathrm{F}$-FDG. The inputs and target for the studies are summarized in Table \ref{tab:cases}. All LR PET images were based on the Siemens ECAT EXACT HR+ scanner. For the first simulation study, the HR PET images were the ``ground-truth" images generated from segmented anatomical templates. For the second simulation study and the clinical study, the HR PET images were based on the Siemens HRRT scanner. Bicubic interpolation was used to resample all input and target images to the same voxel size of 1 mm$\times$1 mm$\times$1 mm with a 256$\times$256$\times$207 grid size. The LR and HR scanner properties are summarized in Table \ref{tab:scanners}.

\begin{table}[!ht]
\caption{Simulation and experimental studies}
\label{tab:cases}
\centering
\begin{tabular}{c|c|c|c}
\hline
\hline
Study index & Study type & LR image (input) & HR image (target) \\
\hline
\hline
1 & Simulation & HR+ PET & True PET \\
\hline
2 & Simulation & HR+ PET & HRRT PET \\
\hline
3 & Clinical & HR+ PET & HRRT PET \\
\hline
\end{tabular}
\end{table}

\begin{table}[!h]
\caption{LR and HR image sources}
\label{tab:scanners}
\centering
\begin{tabular}{l|c|c|c|c}
\hline
\hline
Image type & Scanner & Spatial resolution & Bore diameter & Axial length\\
\hline
\hline
LR & HR+ & 4.3 - 8.3 mm & 562 mm & 155 mm \\
\hline
HR & HRRT & 2.3 - 3.4 mm & 312 mm & 250 mm \\
\hline
\end{tabular}
\end{table}

\subsection{Simulation Setup}
\subsubsection{HR+ PSF Measurement}
\label{sssec:psf}
An experimental measure of the true PSF was made by placing $0.5$ mm diameter sources filled with \fdg inside the HR+ scanner bore $56.2$ cm in diameter and $15.5$ cm in length. The PSF images were reconstructed using ordered subsets expectation maximization (OSEM) with post-smoothing using a Gaussian filter. The PSFs were fitted with Gaussian kernels. We assumed radial and axial symmetry and calculated the PSFs at all other in-between locations as linear combinations of the PSFs measured at the nearest measurement locations. Interpolation weights for the experimental datasets were determined by means of bilinear interpolation over an irregular grid consisting of the quadrilaterals formed by the nearest radial and axial PSF sampling locations from a given point. 
 
\subsubsection{Input Image Generation for Studies 1 and 2}
\label{sssec:inputlrgenfor12}
Realistic simulations were performed using the 3D BrainWeb digital phantom (\url{http://brainweb.bic.mni.mcgill.ca/brainweb/}). 20 distinct atlases with 1 mm isotropic resolution were used to generate a set of ``ground-truth" PET images. The atlases contained the following region labels: gray matter, white matter, blood pool, and cerebrospinal fluid. Static PET images were generated based on a ~1 hour long \fdg scan as described in our earlier paper \cite{Song2019pet}. This ``ground-truth" static PET is referred to as ``true PET" for the rest of the paper. The geometric model of the HR+ scanner was used to generate sinogram data. Noisy data was generated using Poisson deviates of the projected sinograms, a noise model widely accepted in the PET imaging community \cite{Lange1990}. The Poisson deviates were generated with a mean of $10^8$ counts for the full scan duration of 3640 s. The data were then reconstructed using the OSEM algorithm (6 iterations, 16 subsets). The images were subsequently blurred using the measured, spatially-variant PSF to generate the LR PET images. In order to match HR PET image grid size, the LR PET images were interpolated into $256 \times 256 \times 207$ from the HR+ output size of $128 \times 128 \times 64$ using bicubic interpolation. T1-weighted MR images with 1 mm isotropic resolution derived directly from the BrainWeb database were used as HR MR inputs.

\subsubsection{Target Image Generation for Study 1}
\label{sssec:timggenfor1}
In Study 1, our purpose was to train the networks to map LR PET images to ``ground-truth" image domain. The target HR PET images are, therefore, the true PET images as explained previously in section  \ref{sssec:over}.

\subsubsection{Target Image Generation for Study 2}
In Study 2, we trained the networks using simulated HRRT PET images as our target HR images. The geometric model of the HRRT scanner was used to generate sinogram data. Poisson noise realizations were generated for the projected sinograms with a mean of $10^8$ counts for a scan duration of 3640 s. The images were then reconstructed using the OSEM algorithm (6 iterations, 16 subsets). OSEM reconstruction results typically appear grainy due to noise. We, therefore, perform post-filtering with a 2.4 mm full width at half maximum (FWHM) 3D Gaussian filter. Since the intrinsic resolution of the HRRT scanner is in the 2.3-3.4 mm range, this step improves image quality without any appreciable reduction is resolution.

\subsection{Experimental Setup}
Clinical neuroimaging datasets for this paper were obtained from the Alzheimer's Disease Neuroimaging Initiative (ADNI, \url{http://adni.loni.usc.edu/}) database, a public repository containing images and clinical data from 2000+ human datasets. We selected 20 HRRT PET scans and the anatomical T1-weighted MPRAGE MR scans for clinical validation of our method. $10$ of the 20 subjects were from the cognitively normal category. The remaining 10 subjects had mild cognitive impairment. The full scan duration was $30$ minutes ($6 \times 5$-minute frames). The OSEM algorithm (6 iterations, 16 subsets) was used for reconstruction.

\subsubsection{Target Image Generation for Study 3}
As with Study 2 (section \ref{sssec:timggenfor1}), the OSEM-reconstructed HRRT PET images, which were grainy, were post-filtered using a 2.4 mm FWHM 3D Gaussian filter to suppress some of the noise without substantially reducing the image resolution. This led to the target HR PET images for the clinical study.

\subsubsection{Input Image Generation for Study 3}
The LR counterparts of the HRRT images were generated by applying the measured spatially-variant PSF of the HR+ scanner described in section \ref{sssec:psf} to the OSEM-reconstructed HRRT images. While, not directly derived from the HR+ scanner, the use of a measured image-domain PSF ensured parity in terms of spatial resolution with true HR+ images. Rigidly co-registered T1-weighted MR images with $1 \times 1 \times 1$ voxels were used as HR MR inputs. Cross-modality registration was performed using FSL (\url{https://fsl.fmrib.ox.ac.uk})\cite{jenkinson2001,jenkinson2002}.

\subsection{Network Implementation, Training, and Validation}
All networks were implemented on the PyTorch platform. Training was performed using GPU-based acceleration achieved by using an NVIDIA GTX 1080Ti graphic card. An $L_1$ loss function was used for network training. Training was performed using \textit{Adam}, an algorithm for optimizing stochastic objective functions via adaptive estimates of lower order moments \cite{Kingma2014adam}. 

The cohort size (total $20$ subjects) was the same for all three studies. For all studies, training was performed using data from $15$ out of $20$ available subjects to predict a residual image that is an estimated difference between the input LR PET and the ground truth HR PET. The stride and padding for the convolution kernels were both set to $1$. The kernel size was set to $3 \times 3$. All the convolutional layers had $64$ filters except for the last layer, which had only $1$ filter. The batch size was $10$. The learning rate was set to $3 \times 10^{-4}$. The network was trained for $400$ epochs. We validated our results by employing the data from the remaining $5$ subjects.

\subsection{Reference Approaches and Evaluation Metrics}
For the SR PET image, denoted by a vector $\bm{x}\in\mathbb{R}^N$, and the HR MR image, denoted by a vector $\bm{y}\in\mathbb{R}^N$, where $N$ is the number of voxels, the JE penalty is defined as:
\beq
\Phi_{\text{JE}}(\bm{x} | \bm{y}) = -\sum_{i=1}^M \sum_{j=1}^M \delta u \ \delta v \ p(u_i,v_j) \log p(u_i,v_j).
\eeq
Here $\bm{u}\in\mathbb{R}^M$ and $\bm{v}\in\mathbb{R}^M$ are intensity histogram vectors based on the PET and MR images respectively and $M$ is the number of intensity bins. The TV penalty is defined as: 
\beq
\Phi_{\text{TV}}(\bm{x})=(\|\Delta_1 \bm{x}\|_1+\|\Delta_2 \bm{x}\|_1+\|\Delta_3 \bm{x}\|_1),
\eeq
where $\Delta_k$ ($k=1, 2,$ or $3$) are finite difference operators along the three Cartesian coordinate directions.

The evaluation metrics used here are defined below. The true and estimated images are denoted $\bm{x}$ and $\bm{\hat{x}}$ respectively. We use the notation $\mu_{x}$ and $\sigma_{x}$ respectively for the mean and standard deviation of $\bm{x}$.

\subsubsection{Peak Signal-to-Noise Ratio (PSNR)}
The PSNR is the ratio of the maximum signal power to noise power and is defined as:
\beq
\text{PSNR}(\bm{\hat{x}},\bm{x})=20\log_{10}\Big(\frac{\max{(\bm{\hat{x}})}}{\text{RMSE}(\bm{\hat{x}},\bm{x})}\Big),
\eeq
where the root-mean-square error (RMSE) is defined as:
\beq
\text{RMSE}(\bm{\hat{x}},\bm{x})=\sqrt{\frac{1}{N}\sum_k(\hat{x}_k-x_k)^2}.
\eeq
\subsubsection{Structural Similarity Index (SSIM)}
The SSIM is a well-accepted measure of perceived image quality and is defined as:
\beq
\text{SSIM}(\bm{\hat{x}},\bm{x})=\frac{(2\mu_x\mu_{\hat{x}}+c_1)(2\sigma_{x\hat{x}}+c_2)}{(\mu_x^2+\mu_{\hat{x}}^2+c_1)(\sigma_x^2+\sigma_{\hat{x}}^2+c_2)}.
\eeq
Here $c_1$ and $c_2$ are parameters stabilizing the division.

\section{Results}
\label{sec:results}
\subsection{Simulation Results: Study 1}
\begin{figure*}[!ht]
\centering
\includegraphics[width=\linewidth]{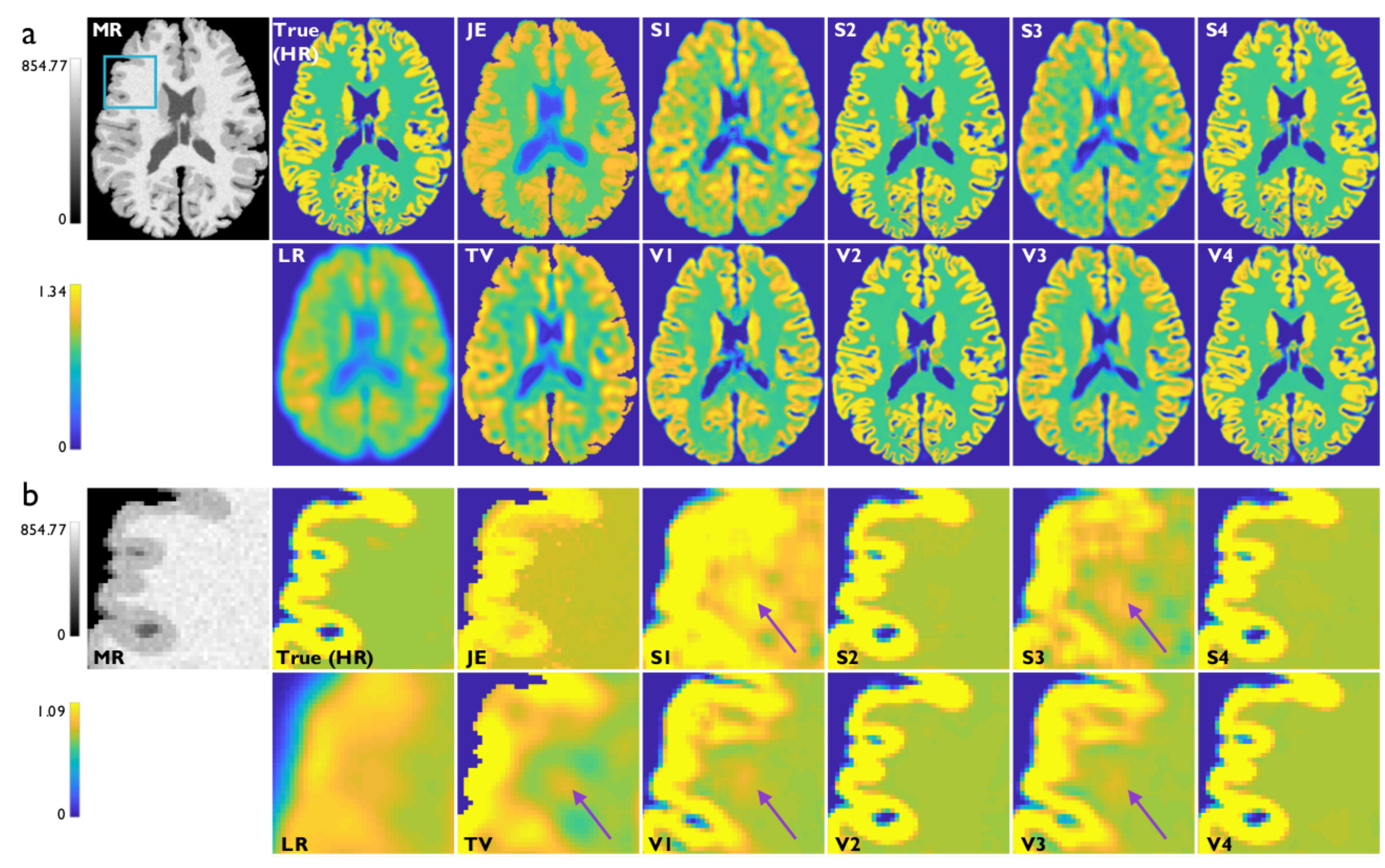}
\caption{Simulation results from the validation set: Study 1. (a) Transverse slices from the T1-weighted HR MR image, true PET image (also the HR image for this case), LR PET image (HR+ scanner), JE-penalized deconvolution result, TV-penalized deconvolution result, and the SR outputs from the following CNNs: S1, V1, S2, V2, S3, V3, S4, and V4. The blue box on the MR image indicates the region that is magnified for closer inspection. (b) The corresponding magnified subimages. Purple arrows indicate areas in the white-matter background region where prominent noise-induced artifacts arise for TV and for the CNNs without anatomical inputs, namely S1, V1, S3, and V4.}
\label{fig:bw_pgt}
\end{figure*}

Fig. \ref{fig:bw_pgt}a showcases results from Study 1 --- transverse slices from the HR MR, HR PET (same as the true PET in this case), LR PET and the SR imaging results from the two deconvolution methods and eight CNNs described in Table \ref{tab:cnn}. The images are for a given subject from the validation dataset. Magnified subimages in Fig. \ref{fig:bw_pgt}b highlight artifacts/inaccuracies indicated by purple arrows that are observed in all the techniques that lack anatomical guidance, namely TV, S1, V1, S3, and V3. Comparison of the subimage pairs (S1, S3) and (V1, V3) illustrates that, in the absence a anatomical information, spatial information greatly enhanced image quality. Comparison of the subimage pairs (S1, V1) and (S3, V3) also shows that the addition of more convolutional layers (increased network depth) is also very effective in the absence of anatomical information. Since JE incorporates anatomical information, it shows better edge recovery than TV. The CNN results show better gray-to-white contrast than both JE and TV. 

The PSNR and SSIM of the different methods are tabulated in Table \ref{tab:study1}. In terms of PSNR, the supervised CNN networks outperform the classical approaches by a wide margin. The networks that have anatomical guidance, S2, S4, V2, and V4 have better performance on both PSNR and SSIM than the networks S1, V1, S3, and V3, which lack anatomical guidance. The PSNR and SSIM figures for S1 vs. S3 shows that these metrics increase noticeably with spatial information for the shallow case.

\begin{table*}[!ht]
\caption{Study 1: Performance comparison}
\label{tab:study1}
\centering
\begin{tabular}{l|c|c|c|c|c|c|c|c|c|c|c|c}
\hline
\hline
Metric &  Reference & LR & TV &  JE  &  S1 & V1 & S2 & V2 & S3 & V3 & S4 & V4 \\
\hline
\hline
PSNR &  True &  21.47 &  22.35 &  22.37 &  26.86 &  27.71 &  37.17 &  37.61 &  27.17 &  27.98 &  37.27 &  \textbf{37.93}\\
\hline
SSIM &  True &  0.74 &  0.85 &  0.86 &  0.74 &  0.86 &  0.97 & 	0.97  &  0.83 &  0.87 &  0.97 &  \textbf{0.98} \\
\hline
\end{tabular}
\end{table*}

\subsection{Simulation Results: Study 2}
\begin{figure*}[!ht]
\centering
\includegraphics[width=\linewidth]{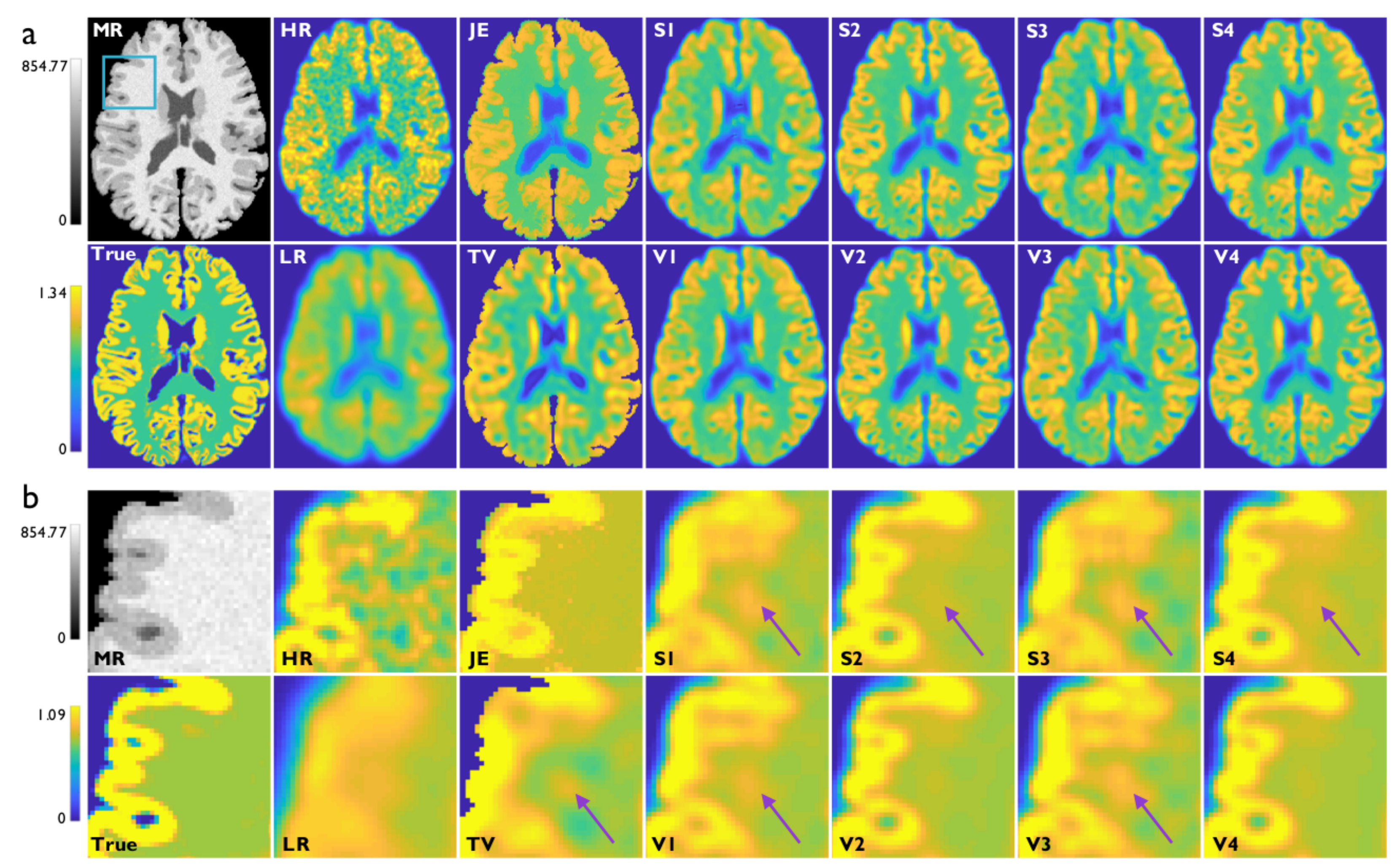}
\caption{Simulation results from the validation set: Study 2. (a) Transverse slices from the T1-weighted HR MR image, true PET image, HR PET image (HRRT scanner), LR PET image (HR+ scanner), JE-penalized deconvolution result, TV-penalized deconvolution result, and the SR outputs from the following CNNs: S1, V1, S2, V2, S3, V3, S4, and V4. The blue box on the MR image indicates the region that is magnified for closer inspection. (b) The corresponding magnified subimages. Purple arrows indicate areas in the white-matter background region where prominent noise-induced artifacts arise. These artifacts are the least prominent for V2 and V4 --- very deep CNNs with anatomical inputs.}
\label{fig:bw_hrrtgt}
\end{figure*}

Fig. \ref{fig:bw_hrrtgt}a showcases results from Study 2 --- transverse slices from the HR MR, true PET, HR PET, LR PET and the SR imaging results from the two deconvolution methods and eight CNNs described in Table \ref{tab:cnn}. The images are for a given subject from the validation dataset. As with Study 1, the CNNs with MR-based anatomical inputs (S2, V2, S4, and V4) still produce the best results. However, since the target image is now a corrupt image with diminished structural similarity with the MR image, the margin of gain from using anatomical information is now reduced. Magnified subimages in Fig. \ref{fig:bw_hrrtgt}b highlight artifacts/inaccuracies indicated by purple arrows that are more preponderant in this study compared to Study 1. Interestingly, for this more challenging problem, the deeper networks (V2 and V4) have reduced background noise variations than their shallower counterparts (S2 and S4), as indicated by purple arrows in the latter. The JE and TV images, which are unsupervised, are the same as those showcased in Fig. \ref{fig:bw_pgt}. 

The PSNR and SSIM of the different methods are tabulated in Table \ref{tab:study2}. An additional goal for this study was to understand the variability in results that could be anticipated when imperfect HR images were used for training. We, therefore, computed two sets of PSNR and SSIM measures: one with respect to the target HR PET and another with respect to the true PET. Our results show that there is an overall reduction in performance when the true PET is used as the reference. This is expected with the HR PET used for training deviates substantially from the true PET. But a key observation here is that the CNNs exhibit consistent relative levels of accuracy for the two reference images. As with Study 1, anatomically-guided networks showed better performance than the non-anatomically guided networks and the two classical methods.

\begin{table*}[!ht]
\caption{Study 2: Performance comparison}
\label{tab:study2}
\centering
\begin{tabular}{l|c|c|c|c|c|c|c|c|c|c|c|c}
\hline
\hline
Metric &  Reference  & LR & TV & JE & S1 & V1 & S2 & V2 & S3 & V3 & S4 & V4 \\
\hline
\hline
PSNR &  HRRT &  27.83 &  26.68 &  27.48 &  35.11 &  35.69 &  38.37 &  38.48 &  35.48 &  35.92 & 38.44 & \textbf{38.69} \\
\hline
PSNR &  True & 21.47 & 22.35 & 22.37 & 22.69 & 23.06 &  22.46 &  23.92 & 23.07 & 23.48 & 23.78 & \textbf{24.29} \\
\hline
SSIM &  HRRT & 0.76 & 0.83 & 0.82 & 0.82 & 0.80 & 0.88 & 0.88 & 0.85 & 0.82 & 0.89 & \textbf{0.90} \\
\hline
SSIM &  True & 0.74 & 0.85 & 0.86 & 0.77 & 0.75 & 0.86 & 0.86 & 0.79 & 0.77 & 0.86 & \textbf{0.87} \\
\hline
\end{tabular}
\end{table*}

\subsection{Experimental Results: Study 3}
\begin{figure*}[!t]
\centering
\includegraphics[width=\linewidth]{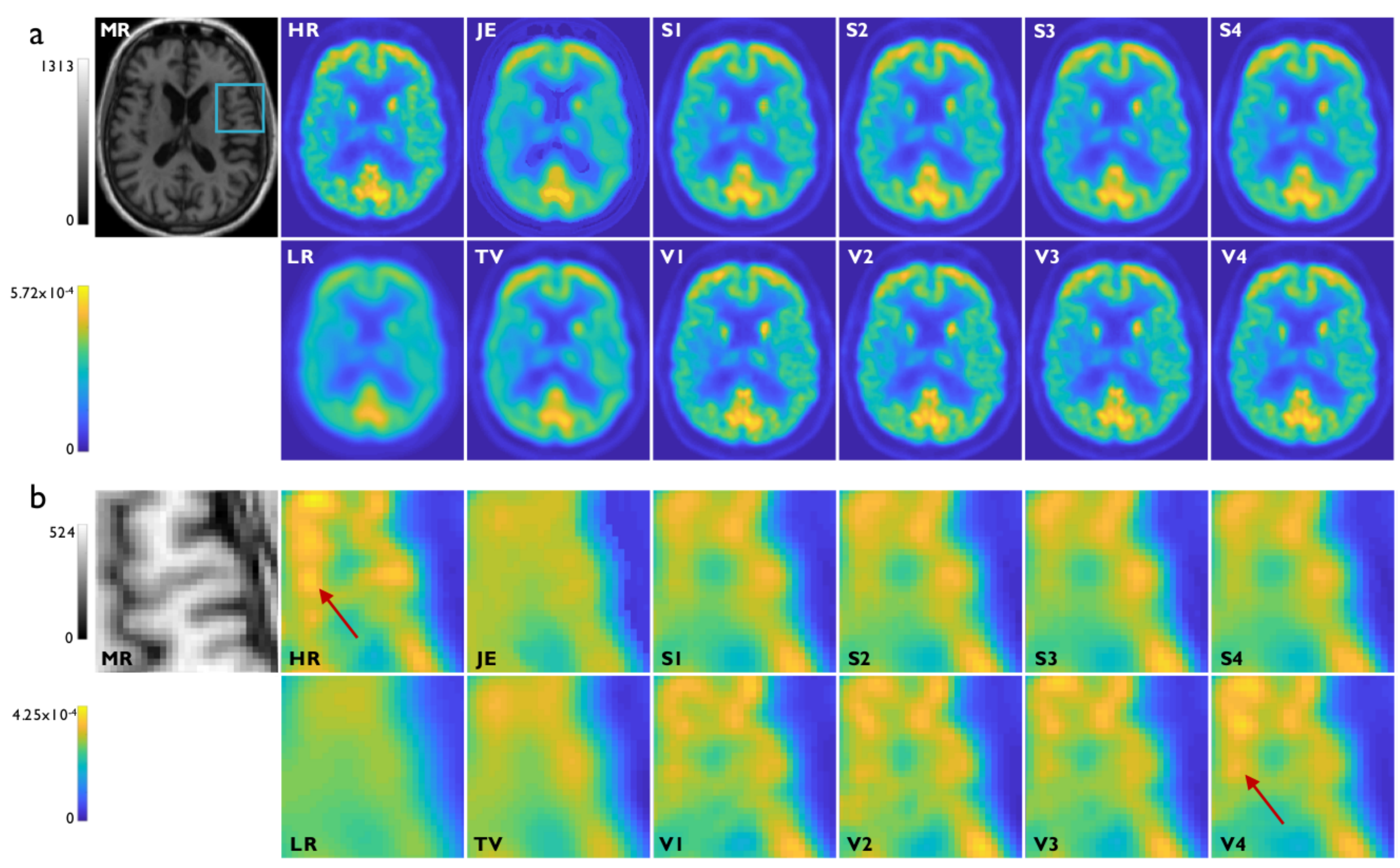}
\caption{Clinical results from the validation set: Study 3. (a) Transverse slices from the T1-weighted HR MR image, HR PET image (HRRT scanner), LR PET image (HR+ scanner), JE-penalized deconvolution result, TV-penalized deconvolution result, and the SR outputs from the following CNNs: S1, V1, S2, V2, S3, V3, S4, and V4. The blue box on the MR image indicates the region that is magnified for closer inspection. (b) The corresponding magnified subimages. The very deep CNNs (V1, V2, V3, and V4) yield sharper images than their shallow counterparts (S1, S2, S3, and S4). The red arrows points to bright gray matter areas  in the HR image that are recovered with the highest contrast in V4.}
\label{fig:hrrt}
\end{figure*}

Fig. \ref{fig:hrrt}a showcases results from Study 3 --- the HR MR, HR PET, LR PET and the SR imaging results from the two deconvolution methods and eight CNNs described in Table \ref{tab:cnn}. For this study, the deeper networks (V1, V2, V3, and V4) produced visually sharper images than their shallower counterparts (S1, S2, S3, and S4). This is clearly evident from the magnified subimages in Fig. \ref{fig:hrrt}b. This is consistent with our previous observation that, in the absence of a strong contribution of the anatomical inputs, the extra layers lead to a stronger margin of improvement. That said, V4, which is deeper and uses anatomical information, led to the highest levels of gray matter contrast as highlighted by red arrows.  

The PSNR, SSIM of the different methods are tabulated in Table \ref{tab:study3}. As displayed in the table, V4 continues to exhibit the best performance in terms of PSNR and SSIM. As in Studies 1 and 2, all CNN based methods outperformed TV and JE. 

\begin{table*}[!ht]
\caption{Study 3: Performance comparison}
\label{tab:study3}
\centering
\begin{tabular}{l|c|c|c|c|c|c|c|c|c|c|c|c}
\hline
\hline
Metric & Reference & LR & TV & JE & S1 & V1 & S2 & V2 & S3 & V3 & S4 & V4 \\
\hline
\hline
PSNR &  HRRT &  29.93 &  34.98 &  33.32 &  35.75 &  36.93 &  36.00 & 37.27 &  36.29 &  37.42 &  36.46  &  \textbf{38.02} \\
\hline
SSIM &  HRRT &  0.86 &  0.93 &  0.88 &  0.94 &  0.95 &  0.95 &  0.95 &  0.95 &  0.95 &  0.95 &  \textbf{0.96} \\
\hline
\end{tabular}
\end{table*}

\section{Discussion}
\label{sec:discussion}

Overall, our results indicate that CNNs vastly outperform penalized deconvolution at the SR imaging task. Among different CNN architectures, relative performance depends on the problem at hand. Our simulation and clinical studies all agree that deep CNNs outperform shallow CNNs and that the additional channels contribute to improving overall performance. The relative importance of anatomical and spatial input channel depends on the underlying structural similarity between the HR MR and the true PET.

It should be noted that the utilization of anatomical images in conjunction with functional images required co-registration. To ensure robustness, a well-tested standardized registration tool was used for the task. Since the registration is intra-subject, rigid registration based on mutual information suffices for this application. 

A limitation of the CNNs designed and validated in this paper is that they are specific to inputs with noise and blur levels similar to those used in training. In other words, they lack portability. Even for a given LR-HR scanner pair, such as the HR+ and HRRT, the SR performance is expected to drop when the LR inputs are based on a much lower or higher tracer dose than the datasets used for training. Our future work will characterize the performance of these networks when input noise levels are varied. One remedial approach to address noise sensitivity is the use of transfer learning for easy retraining of the networks with a much smaller training dataset containing LR inputs with altered SNR. Another, more sophisticated strategy is to use the CNN output as a prior in reconstruction. We have previously used this approach and produced promising results for denoising \cite{Kim2018pen} and anticipate that it will work for deblurring and SR by extension. 

Another limitation of CNN-based SR, and perhaps the most significant one, is that it relies on supervised learning and, therefore, requires paired LR PET and HR PET images for training. This requirement is easy to address while training using simulated datasets. But paired LR and HR clinical scans are rare. To address this limitation, we are currently exploring self-supervised learning strategies based on adversarial training of generative adversarial networks, that circumvent the need for paired training inputs.

\section{Conclusion}
\label{sec:conclusion}

We have designed, implemented, and validated a family of CNN-based SR PET imaging techniques. To facilitate the resolution recovery process, we incorporated both anatomical and spatial information. In the studies presented here, the anatomical information was provided as an HR MR image. So as to easily provide spatial information, we supplied patches containing the radial and axial coordinates of each voxel as additional input channels. This strategy is well-consistent with standard CNN multi-channel input formats and, therefore, convenient to implement.  Both simulation and clinical studies showed that the CNNs greatly outperform penalized deconvolution both qualitatively (e.g., edge and contrast recovery) and quantitatively (as indicated by PSNR and SSIM). 

As future work, we will develop new networks based on self-supervised learning that will enable us to circumvent the need for paired training datasets. We will also explore avenues to ensure applicability of this method to super-resolve inputs with noise levels different from those in the training data. From an applications perspective, we are interested in using this technique to super-resolve PET images of tau tangles, a neuropathological hallmark of Alzheimer's disease, with the goal of developing sensitive image-based biomarkers for tau. 




\bibliographystyle{unsrt}  

\bibliography{Song_SR_refs}

\begin{thebibliography}{10}

\bibitem{Sotoudeh_jmri}
H.~Sotoudeh, A.~Sharma, K.~J. Fowler, J.~McConathy, and F.~Dehdashti.
\newblock {{C}linical application of {P}{E}{T}/{M}{R}{I} in oncology}.
\newblock {\em J. Magn. Reson. Imaging}, 44(2):265--276, 08 2016.

\bibitem{Catana_jnm}
C.~Catana, A.~Drzezga, W.~D. Heiss, and B.~R. Rosen.
\newblock {{P}{E}{T}/{M}{R}{I} for neurologic applications}.
\newblock {\em J. Nucl. Med.}, 53(12):1916--1925, Dec 2012.

\bibitem{Salata2017role}
B.~M. Salata and P.~Singh.
\newblock {{R}ole of {C}ardiac {P}{E}{T} in {C}linical {P}ractice}.
\newblock {\em Curr. Treat. Options Cardiovasc. Med.}, 19(12):93, Nov 2017.

\bibitem{Hess2016pet}
S.~Hess, A.~Alavi, and S.~Basu.
\newblock {{P}{E}{T}-{B}ased {P}ersonalized {M}anagement of {I}nfectious and
  {I}nflammatory {D}isorders}.
\newblock {\em PET Clin.}, 11(3):351--361, Jul 2016.

\bibitem{leahy2000stat}
Richard~M Leahy and Jinyi Qi.
\newblock Statistical approaches in quantitative positron emission tomography.
\newblock {\em Stat. Comput.}, 10(2):147--165, Apr 2000.

\bibitem{dutta2013quant}
J.~Dutta, S.~Ahn, and Q.~Li.
\newblock {{Q}uantitative statistical methods for image quality assessment}.
\newblock {\em Theranostics}, 3(10):741--756, Oct 2013.

\bibitem{dutta2013non}
J.~Dutta, R.~M. Leahy, and Q.~Li.
\newblock {{N}on-local means denoising of dynamic {P}{E}{T} images}.
\newblock {\em PLoS ONE}, 8(12):e81390, Dec 2013.

\bibitem{qi2000res}
J.~Qi and R.~M. Leahy.
\newblock {{R}esolution and noise properties of {M}{A}{P} reconstruction for
  fully 3-{D} {P}{E}{T}}.
\newblock {\em IEEE Trans. Med. Imaging}, 19(5):493--506, May 2000.

\bibitem{rousset1998}
O.~G. Rousset, Y.~Ma, and A.~C. Evans.
\newblock {{C}orrection for partial volume effects in {P}{E}{T}: principle and
  validation}.
\newblock {\em J. Nucl. Med.}, 39(5):904--911, May 1998.

\bibitem{reader2003algorithm}
Andrew~J Reader, Peter~J Julyan, Heather Williams, David~L Hastings, and Jamal
  Zweit.
\newblock {EM} algorithm system modeling by image-space techniques for {PET}
  reconstruction.
\newblock {\em IEEE Trans. Nucl. Sci.}, 50(5):1392--1397, Oct 2003.

\bibitem{alessio2006}
A.~M. Alessio, P.~E. Kinahan, and T.~K. Lewellen.
\newblock {{M}odeling and incorporation of system response functions in 3-{D}
  whole body {P}{E}{T}}.
\newblock {\em IEEE Trans. Med. Imaging}, 25(7):828--837, Jul 2006.

\bibitem{panin2006}
V.~Y. Panin, F.~Kehren, C.~Michel, and M.~Casey.
\newblock {{F}ully 3-{D} {P}{E}{T} reconstruction with system matrix derived
  from point source measurements}.
\newblock {\em IEEE Trans. Med. Imaging}, 25(7):907--921, Jul 2006.

\bibitem{leahy1991}
Richard Leahy and X~Yan.
\newblock Incorporation of anatomical {MR} data for improved functional imaging
  with {PET}.
\newblock In {\em Inf. Process. Med. Imaging.}, volume 511, pages 105--120.
  Springer, 1991.

\bibitem{comtat2002}
C.~Comtat, P.~E. Kinahan, J.~A. Fessler, T.~Beyer, D.~W. Townsend, M.~Defrise,
  and C.~Michel.
\newblock {{C}linically feasible reconstruction of 3{D} whole-body
  {P}{E}{T}/{C}{T} data using blurred anatomical labels}.
\newblock {\em Phys. Med. Biol.}, 47(1):1--20, Jan 2002.

\bibitem{bowsher2004utilizing}
James~E Bowsher, Hong Yuan, Laurence~W Hedlund, Timothy~G Turkington, Gamal
  Akabani, Alexandra Badea, William~C Kurylo, C~Ted Wheeler, Gary~P Cofer,
  Mark~W Dewhirst, et~al.
\newblock Utilizing {MRI} information to estimate {F18-FDG} distributions in
  rat flank tumors.
\newblock In {\em IEEE Nucl. Sci. Symp. Conf. Rec.}, volume~4, pages
  2488--2492. IEEE, 2004.

\bibitem{baete2004}
K.~Baete, J.~Nuyts, W.~Van~Paesschen, P.~Suetens, and P.~Dupont.
\newblock {{A}natomical-based {F}{D}{G}-{P}{E}{T} reconstruction for the
  detection of hypo-metabolic regions in epilepsy}.
\newblock {\em IEEE Trans. Med. Imaging}, 23(4):510--519, Apr 2004.

\bibitem{bataille2007brain}
F~Bataille, C~Comtat, S~Jan, FC~Sureau, and R~Trebossen.
\newblock Brain {PET} partial-volume compensation using blurred anatomical
  labels.
\newblock {\em IEEE Trans. Nucl. Sci.}, 54(5):1606--1615, Apr 2007.

\bibitem{pedemonte20114D}
S.~Pedemonte, A.~Bousse, B.~F. Hutton, S.~Arridge, and S.~Ourselin.
\newblock {4-{D} generative model for {P}{E}{T}/{M}{R}{I} reconstruction}.
\newblock {\em Med. Image Comput. Comput. Assist. Interv.}, 14(Pt 1):581--588,
  2011.

\bibitem{somayajula2011}
S.~Somayajula, C.~Panagiotou, A.~Rangarajan, Q.~Li, S.~R. Arridge, and R.~M.
  Leahy.
\newblock {{P}{E}{T} image reconstruction using information theoretic
  anatomical priors}.
\newblock {\em IEEE Trans. Med. Imaging}, 30(3):537--549, Mar 2011.

\bibitem{wang2012pen}
G.~Wang and J.~Qi.
\newblock {{P}enalized likelihood {P}{E}{T} image reconstruction using
  patch-based edge-preserving regularization}.
\newblock {\em IEEE Trans. Med. Imaging}, 31(12):2194--2204, Dec 2012.

\bibitem{kim2015dynamic}
K.~Kim, Y.~D. Son, Y.~Bresler, Z.~H. Cho, J.~B. Ra, and J.~C. Ye.
\newblock {{D}ynamic {P}{E}{T} reconstruction using temporal patch-based low
  rank penalty for {R}{O}{I}-based brain kinetic analysis}.
\newblock {\em Phys. Med. Biol.}, 60(5):2019--2046, Mar 2015.

\bibitem{wang2015edge}
G.~Wang and J.~Qi.
\newblock {{E}dge-preserving {P}{E}{T} image reconstruction using trust
  optimization transfer}.
\newblock {\em IEEE Trans. Med. Imaging}, 34(4):930--939, Apr 2015.

\bibitem{meltzer1990}
C.~C. Meltzer, J.~P. Leal, H.~S. Mayberg, H.~N. Wagner, and J.~J. Frost.
\newblock {{C}orrection of {P}{E}{T} data for partial volume effects in human
  cerebral cortex by {M}{R} imaging}.
\newblock {\em J. Comput. Assist. Tomogr.}, 14(4):561--570, Jul-Aug 1990.

\bibitem{mullergartner1992}
H.~W. M\"uller-G\"artner, J.~M. Links, J.~L. Prince, R.~N. Bryan, E.~McVeigh,
  J.~P. Leal, C.~Davatzikos, and J.~J. Frost.
\newblock {M}easurement of radiotracer concentration in brain gray matter using
  positron emission tomography: {M}{R}{I}-based correction for partial volume
  effects.
\newblock {\em J. Cereb. Blood Flow Metab.}, 12(4):571--583, Jul 1992.

\bibitem{soret2007}
M.~Soret, S.~L. Bacharach, and I.~Buvat.
\newblock {{P}artial-volume effect in {P}{E}{T} tumor imaging}.
\newblock {\em J. Nucl. Med.}, 48(6):932--945, Jun 2007.

\bibitem{thomas2011}
B.~A. Thomas, K.~Erlandsson, M.~Modat, L.~Thurfjell, R.~Vandenberghe,
  S.~Ourselin, and B.~F. Hutton.
\newblock {{T}he importance of appropriate partial volume correction for
  {P}{E}{T} quantification in {A}lzheimer's disease}.
\newblock {\em Eur. J. Nucl. Med. Mol. Imaging}, 38(6):1104--1119, Jun 2011.

\bibitem{bousse2012}
Alexandre Bousse, Stefano Pedemonte, Benjamin~A Thomas, Kjell Erlandsson,
  S{\'e}bastien Ourselin, Simon Arridge, and Brian~F Hutton.
\newblock Markov random field and {G}aussian mixture for segmented {MRI}-based
  partial volume correction in {PET}.
\newblock {\em Phys. Med. Biol.}, 57(20):6681, 2012.

\bibitem{vancittert1931}
P~H Van~Cittert.
\newblock Zum {E}influ\ss\ der {S}paltbreite auf die {I}ntensit\"atsverteilung
  in {S}pektrallinien {II}.
\newblock {\em Z. Phys.}, 69(5-6):298--308, May 1931.

\bibitem{richardson1972bayesian}
William~Hadley Richardson.
\newblock Bayesian-based iterative method of image restoration.
\newblock {\em J. Opt. Soc. Am.}, 62(1):55--59, Jan 1972.

\bibitem{lucy1974iterative}
Leon~B Lucy.
\newblock An iterative technique for the rectification of observed
  distributions.
\newblock {\em Astron. J.}, 79:745, Jun 1974.

\bibitem{yan2015}
J.~Yan, J.~C. Lim, and D.~W. Townsend.
\newblock {{M}{R}{I}-guided brain {P}{E}{T} image filtering and partial volume
  correction}.
\newblock {\em Phys. Med. Biol.}, 60(3):961--976, Feb 2015.

\bibitem{Song2019pet}
T-A Song, F~Yang, SR~Chowdhury, K~Kim, KA~Johnson, G~El~Fakhri, Q~Li, and
  J~Dutta.
\newblock {PET} image deblurring and super-resolution with an {MR}-based joint
  entropy prior.
\newblock {\em IEEE Trans Comput Imaging}, 2019.

\bibitem{Nasrollahi2012super}
K~Nasrollahi and TB~Moeslund.
\newblock Super-resolution: A comprehensive survey.
\newblock {\em Mach. Vis. Appl.}, 25:1423--68, 2014.

\bibitem{Wallach2012super}
D~Wallach, F~Lamare, G~Kontaxakis, and D~Visvikis.
\newblock Super-resolution in respiratory synchronized positron emission
  tomography.
\newblock {\em IEEE Trans. Med. Imaging}, 31(2):438--448, 2012.

\bibitem{Glasner2009}
Glasner D, Bagon S, and Irani M.
\newblock Super-resolution from a single image.
\newblock {\em Proc. IEEE Int. Conf. Comput. Vis.}, pages 349--56, 2009.

\bibitem{Kim2010SingleImageSR}
Kwang~In Kim and Younghee Kwon.
\newblock Single-image super-resolution using sparse regression and natural
  image prior.
\newblock {\em IEEE Trans. Pattern Anal. Mach. Intell.}, 32(6):1127--1133, June
  2010.

\bibitem{Yang2010image}
Yang J, Wright J, Huang TS, and Ma~Y.
\newblock Image super-resolution via sparse representation.
\newblock {\em IEEE Trans. Image Process.}, 19(11):2861--2873, November 2010.

\bibitem{Timofte2013}
Timofte R, De~Smet V, and Van~Gool L.
\newblock Anchored neighborhood regression for fast example-based
  super-resolution.
\newblock {\em Proc. IEEE Int. Conf. Comput. Vis.}, pages 1920--1927, 2013.

\bibitem{Yang2013fast}
Yang J, Lin Z, and Cohen S.
\newblock Fast image super-resolution based on in-place example regression.
\newblock {\em Proc. IEEE Int. Conf. Comput. Vis.}, pages 1059--1066, 2013.

\bibitem{Jia2013}
Jia K, Wang X, and Tang X.
\newblock Image transformation based on learning dictionaries across image
  spaces.
\newblock {\em IEEE Trans. Pattern Anal. Mach. Intell.}, 35(11):367--380,
  February 2013.

\bibitem{Schulter2013}
Schulter S, Leistner C, and Bischof H.
\newblock Fast and accurate image upscaling with super-resolution forests.
\newblock {\em Proc. IEEE Int. Conf. Comput. Vis.}, pages 3791--99, 2015.

\bibitem{Dong2016image}
Chao Dong, Chen~Change Loy, Kaiming He, and Xiaoou Tang.
\newblock Image super-resolution using deep convolutional networks.
\newblock {\em IEEE Trans. Pattern Anal. Mach. Intell.}, 38(2):295--307, 2016.

\bibitem{Kim2016accurate}
Jiwon Kim, Jung Kwon~Lee, and Kyoung Mu~Lee.
\newblock Accurate image super-resolution using very deep convolutional
  networks.
\newblock In {\em Proc. IEEE Comput. Soc. Conf. Comput. Vis. Pattern
  Recognit.}, pages 1646--1654, 2016.

\bibitem{He2009deep}
K~He, X~Zhang, S~Ren, and J~Sun.
\newblock Deep residual learning for image recognition.
\newblock {\em Proc. IEEE Int. Conf. Comput. Vis.}, pages 349--56, 2009.

\bibitem{Ledig2017photo}
Christian Ledig, Lucas Theis, Ferenc Huszar, Jose Caballero, Andrew Cunningham,
  Alejandro Acosta, Andrew Aitken, Alykhan Tejani, Johannes Totz, Zehan Wang,
  and Wenzhe Shi.
\newblock Photo-realistic single image super-resolution using a generative
  adversarial network.
\newblock {\em Proc. IEEE Int. Conf. Comput. Vis.}, pages 105--114, 2017.

\bibitem{Dutta2015pet}
Joyita Dutta, Georges El~Fakhri, Xuping Zhu, and Quanzheng Li.
\newblock {PET} point spread function modeling and image deblurring using a
  {PET/MRI} joint entropy prior.
\newblock In {\em Proc. IEEE Int. Symp. Biomed. Imaging}, pages 1423--1426.
  IEEE, 2015.

\bibitem{Song2018super}
T-A Song, SR~Chowdhury, K~Kim, K~Gong, G~El~Fakhri, Q~Li, and J~Dutta.
\newblock Super-resolution {PET} using a very deep convolutional neural
  network.
\newblock In {\em Proc. IEEE Nucl. Sci. Symp. Med. Imag. Conf.} IEEE, 2018.

\bibitem{cloquet2010}
C.~Cloquet, F.~C. Sureau, M.~Defrise, G.~Van~Simaeys, N.~Trotta, and
  S.~Goldman.
\newblock {{N}on-{G}aussian space-variant resolution modelling for list-mode
  reconstruction}.
\newblock {\em Phys. Med. Biol.}, 55(17):5045--5066, Sep 2010.

\bibitem{Gu2018RecentAI}
Jiuxiang Gu, Zhenhua Wang, Jason Kuen, Lianyang Ma, Amir Shahroudy, Bing Shuai,
  Ting Liu, Xingxing Wang, Gang Wang, Jianfei Cai, and Tsuhan Chen.
\newblock Recent advances in convolutional neural networks.
\newblock {\em Pattern Recognition}, 77:354--377, 2018.

\bibitem{LeCun1998EfficientB}
Yann LeCun, L{\'e}on Bottou, Genevieve~B. Orr, and Klaus-Robert M{\"u}ller.
\newblock Efficient backprop.
\newblock In {\em Neural Networks: Tricks of the Trade}, 1998.

\bibitem{Lim2017EnhancedDR}
Bee Lim, Sanghyun Son, Heewon Kim, Seungjun Nah, and Kyoung~Mu Lee.
\newblock Enhanced deep residual networks for single image super-resolution.
\newblock {\em Proc. IEEE Comput. Soc. Conf. Comput. Vis. Pattern Recognit.},
  pages 1132--1140, 2017.

\bibitem{Lange1990}
K.~Lange.
\newblock {{C}onvergence of {E}{M} image reconstruction algorithms with {G}ibbs
  smoothing}.
\newblock {\em IEEE Trans. Med. Imaging}, 9(4):439--446, 1990.

\bibitem{jenkinson2001}
M.~Jenkinson and S.~Smith.
\newblock {{A} global optimisation method for robust affine registration of
  brain images}.
\newblock {\em Med. Image Anal.}, 5(2):143--156, Jun 2001.

\bibitem{jenkinson2002}
M.~Jenkinson, P.~Bannister, M.~Brady, and S.~Smith.
\newblock {{I}mproved optimization for the robust and accurate linear
  registration and motion correction of brain images}.
\newblock {\em Neuroimage}, 17(2):825--841, Oct 2002.

\bibitem{Kingma2014adam}
DP~Kingma and J~Ba.
\newblock Adam: A method for stochastic optimization.
\newblock {\em arXiv preprint arXiv:1412.6980}, 2014.

\bibitem{Kim2018pen}
K.~Kim, D.~Wu, K.~Gong, J.~Dutta, J.~H. Kim, Y.~D. Son, H.~K. Kim,
  G.~El~Fakhri, and Q.~Li.
\newblock {{P}enalized {P}{E}{T} {R}econstruction {U}sing {D}eep {L}earning
  {P}rior and {L}ocal {L}inear {F}itting}.
\newblock {\em IEEE Trans. Med. Imaging}, 37(6):1478--1487, 06 2018.

\end{thebibliography}
%
%
%
%

\end{document}